# Hyperuniformity in active matter

Commentary by

**Hartmut Löwen**

Institut für Theoretische Physik II: Weiche Materie, Department of Physics, Heinrich-Heine-University Düsseldorf, Universitätsstraße 1, 40225 Düsseldorf, Germany

**Erdal C. Oğuz**

Key Laboratory of Soft Matter Physics, Institute of Physics, Chinese Academy of Sciences, Beijing 100190, China

on

**Nonequilibrium strongly hyperuniform fluids of circle active particles with large local density fluctuations**

Q. L. Lei, M. P. Ciamarra, R. Ni, Science Advances 5, eaau7423 (2019)

## Statement of Significance

*Disordered hyperuniform systems are characterized by an anomalously strong suppression of long-wavelength density fluctuations. As a result, these unusual materials behave as though virtually incompressible, even though the particle arrangement itself can remain fully disordered at short range. This peculiar state of matter has been the topic of intense research in physics and mathematics in recent years. Another flourishing research arena is active matter, composed of self-propelled agents that convert energy into motion. While the two fields remained largely disjunct for a long time, the paper by Lei and coworkers (2019) created a link between hyperuniformity and active matter by demonstrating that interacting circle swimmers self-organize into disordered hyperuniform structures. Inspired by this work, many other examples of hyperuniformity in active matter have been identified. We shall describe the impact and outgrowth of this development, and give a brief perspective on further links between hyperuniformity and active matter.*

## Introduction

Structural correlations in classical liquids are typically characterized by the liquid structure factor $S(k)$ [1]. This dimensionless function quantifies autocorrelations of density waves with wave number $k$. If $S(k)$ is large, the probability to find density waves with wave number *k* is high, while if $S(k)$ vanishes there are no such density wave excitations in the spectrum. A typical equilibrium liquid structure factor is shown in Figure 1a. At low *k*, S(*k*) starts at a finite positive value, proportional to the isothermal compressibility. It then exhibits a first main peak at a wave number *k**, which typically scales with the inverse average interparticle distance in the liquid. This reflects the presence of density waves where particle configurations keep this distance. As a function of $k$, $S(k)$ then oscillates and approaches 1 for large $k$. But there are important exceptions:

i) In an *ideal gas*, there are no correlations at all, so $S(k) = 1$ for any *k*.
ii) In a *critical* liquid, the compressibility diverges, hence $S(\mathrm{k} \to 0)$ tends to infinity (also referred to as antihyperuniform).
iii) There is a separate class of liquids which are ***hyperuniform*** which in equilibrium implies that they are incompressible; hence $S(\mathrm{k} \to 0) = 0$.

As a consequence of hyperuniformity, actual particle positions are uniformly distributed in such a way that long-wavelength density fluctuations are suppressed completely, i.e., *S(k)*→0 as *k*→0. Such systems can be further categorized into different classes [2], distinguished by their asymptotic behavior as *k*→0. Assuming $S(k)$ is proportional to $k^{\alpha}$, one can discriminate:

i) *Class I* hyperuniformity if $\alpha > 1$
ii) *Class II* hyperuniformity if $\alpha = 1$
iii) *Class III* hyperuniformity if $0 < \alpha < 1$

Class I represents the strongest form of density-fluctuation suppression, reminiscent of that found in perfect crystals. It also includes the special case of *stealthy hyperuniformity*, in which $S(k)$ vanishes identically for all 0 < *k* < *K* with a finite cutoff wave number $K > 0$, giving rise to unusual material properties [3]. Clearly a double logarithmic plot of $S(k)$ at low $k$ reveals whether hyperuniformity is present at all and, if so, which class is realized, see Figure 1b.

In equilibrium, a paradigmatic example is the *one-component plasma* with a constant fixed background required to stabilize the total system in the thermodynamic limit [1]. The one-component plasma is incompressible due to the underlying oppositely charged background: if the particles are compressed but the background is kept fixed, this needs an infinite amount of Coulomb energy, such that charge density waves are completely suppressed and therefore strictly $S(\mathrm{k} \to 0) = 0$ , which, more technically, is embodied in the Stillinger-Lovett sum rule for the plasma. More generally, in equilibrium, this suppression of long-wavelength fluctuations is not unique to the Coulombic case: any classical fluid of particles interacting via a repulsive power-law potential $v(r) \sim 1/r^s$ with *s*<*d* ("strongly" long-range) exhibits hyperuniformity, while *s*>*d* ("weakly" long-range) interactions do not [4], *d* being the dimension of the system.

Hyperuniformity can also be described within a more coarse-grained *field-theory*. For a scalar field $\phi(\boldsymbol{r})$, with $\boldsymbol{r}$ denoting the position vector, hyperuniformity is characterized by a vanishing spectral density in the long wavelength limit, $\lim_{k\to 0} \tilde{\psi}\,(\boldsymbol{k}) = 0$, where $\tilde{...}$ denotes a Fourier transform. Here, $\psi(\boldsymbol{r}) = \langle\, (\phi(\boldsymbol{r}_1) - \bar{\phi})(\phi(\boldsymbol{r}_2) - \bar{\phi})\,\rangle$ where $\langle ... \rangle$ is the spatial average over the field,$\bar{\phi} = \langle \phi(\boldsymbol{r}) \rangle$ is the mean value of the field and $\boldsymbol{r} = \boldsymbol{r}_2 - \boldsymbol{r}_1$ the distance between the two positions of the field, assuming translational invariance of the system's state. Since the field is already an averaged quantity, this criterion is more coarse-grained than its full particle-based definition involving $S(k) = 0$ but still provides an efficient route to identify hyperuniformity.

The study of hyperuniformity is a flourishing research field: it has been discovered in various materials relevant for physics, biology and materials science (for a review, see [2]; for a recent example, see [5]).

Another research field is that of **active matter** [6]. Sparked by the Vicsek model of flocking in 1995 [7], active matter encompasses systems composed of active agents that internally convert energy into mechanical motion. There are many examples in the macroscopic world (including pedestrians, fish, and birds) and in the mesoscopic world (such as bacteria, microbes, and actin filaments). Self-propelled agents can also be synthesized artificially; typical examples are robots and active Janus-colloids.

The fundamental insight of [8] is that hyperuniformity can be hidden and generated in active matter systems. This points to an important link between hyperuniformity and active matter.

## Hyperuniformity in interacting circle swimmers

Lei and coworkers [8] studied a dense system of circle swimmers obeying overdamped motion for both their translation and orientation in the two-dimensional plane (see Figure 2 for a sketch of the model). The orientation vector performs a constant rotation with a spinning frequency ω, while the particle translation occurs with a constant speed $v_0$ along a circle - i.e., a single isolated particle undergoes a perfect circular motion in the plane. The particles are also subject to non-aligning short-ranged repulsive interactions with their neighbours. The mutual collisions with neighbouring circlers produce a steady state in which the particles avoid most of the collisions and distribute themselves almost uniformly. Computing the corresponding liquid structure factor in this self-organized steady state reveals hyperuniformity within a certain region of parameter space. It is of Class I with an exponent $\alpha = 2$.

This is an important result: here, the activity and the particle interactions together generate hyperuniformity, rather than the interparticle interaction alone as in common equilibrium situations. Hyperuniformity in chiral circle swimmers was later confirmed experimentally by studying the swimming of microalgae in quasi-2d confinement [9], and reproduced within a consistent field-theoretic description [10].

## Hyperuniform particle configurations in active matter

The paper of Lei et al. [8] stimulated the exploration other active-matter systems in which activity, together with the interactions between active agents, produces hyperuniform particle configurations.

These can be broadly classified into two distinct cases: The first involves *long-ranged interactions* such as the Coulomb interaction. Here one can start from equilibrium hyperuniformity of the one-component plasma. If the charged particles are made active - for example, Janus-particles in the plasma that self-propel via light [11] (see also [12]), hyperuniformity is retained even in the presence of activity, again falling into Class I. Again, despite the activity, it is the long-ranged Coulomb interaction between the charged particles with the oppositely fixed background that completely suppresses large wavelength fluctuations.

A similar situation occurs for nematic defects in two-dimensional liquid crystals. If the defect positions are treated as effective “particles”, they interact via long-ranged forces that determine their motion and configuration. In two dimensions, the interaction can be logarithmic in the defect separation, similar to the interaction between parallel charged rods perpendicular to the plane. Again, this generates hyperuniform structures due to the long range of the repulsive interaction. The same picture is expected for active nematics, where hyperuniform defect structures are found due to their long-ranged interactions [13].

The second, and perhaps more interesting case, involves *short-ranged interactions,* where the interplay between subsequent cascades of interactions resp. collisions produces hyperuniformity. Beyond the first discovery of Lei et al [8], more and more examples have since been identified in active matter. These include anti-aligning interactions where particles try to avoid each other [14] and thus stay apart to form hyperuniform configurations. Other important examples are vertex and Voronoi models for confluent tissue of moving cells. If the cell volume is restricted, this constraint produces strong hyperuniformity with relatively high exponents $\alpha$ [15].

## Hyperuniformity in active field theories

Hyperuniform structures can also be studied using coarse-grained field theory, which has provided more general insight how activity ~~to~~ generates hyperuniformity. Three important examples illustrate this point.

First, the basic case of *circle swimmers* considered in Lei et al [8] has been described by field theory by Kuroda and Miyasaki [10]. This comprehensive coherent approach provides an insight into the need for chirality in producing hyperuniform structures - in other words, linear (non-chiral) swimmers do not produce hyperuniformity.

Second, extended Navier-Stokes equations for the particle density have been developed to describe *active turbulence* [16]. Recently, numerical work has shown that structures in this field theory are hyperuniform [17].

The third example concerns active field theories which extend the Cahn-Hilliard model of phase separation to describe MIPS (*motility-induced phase separation*). Since passive Cahn-Hilliard dynamics already produces hyperuniform structures (under certain conditions), the question arises whether this behaviour is fundamentally altered, or remains universal, when the theory is extended to include activity. For a large class of couplings, it has indeed been confirmed that hyperuniformity is persistent and occurs also in motility-induced phase separation [18, 19].

## **Conclusions**

In conclusion, activity is now recognized as an important ingredient for generating disordered hyperuniform structures. This was first exemplified in the study by Lei et al. [8], and has since been found in a diversity of other active matter systems. Looking ahead, one can expect even stronger hyperuniform structures by using machine learning techniques to design particle interactions. One can also equip particles with artificial intelligence [20] to program them to avoid each other. This may create very strong hyperuniformity with large alpha exponents, or even stealthy hyperuniformity.

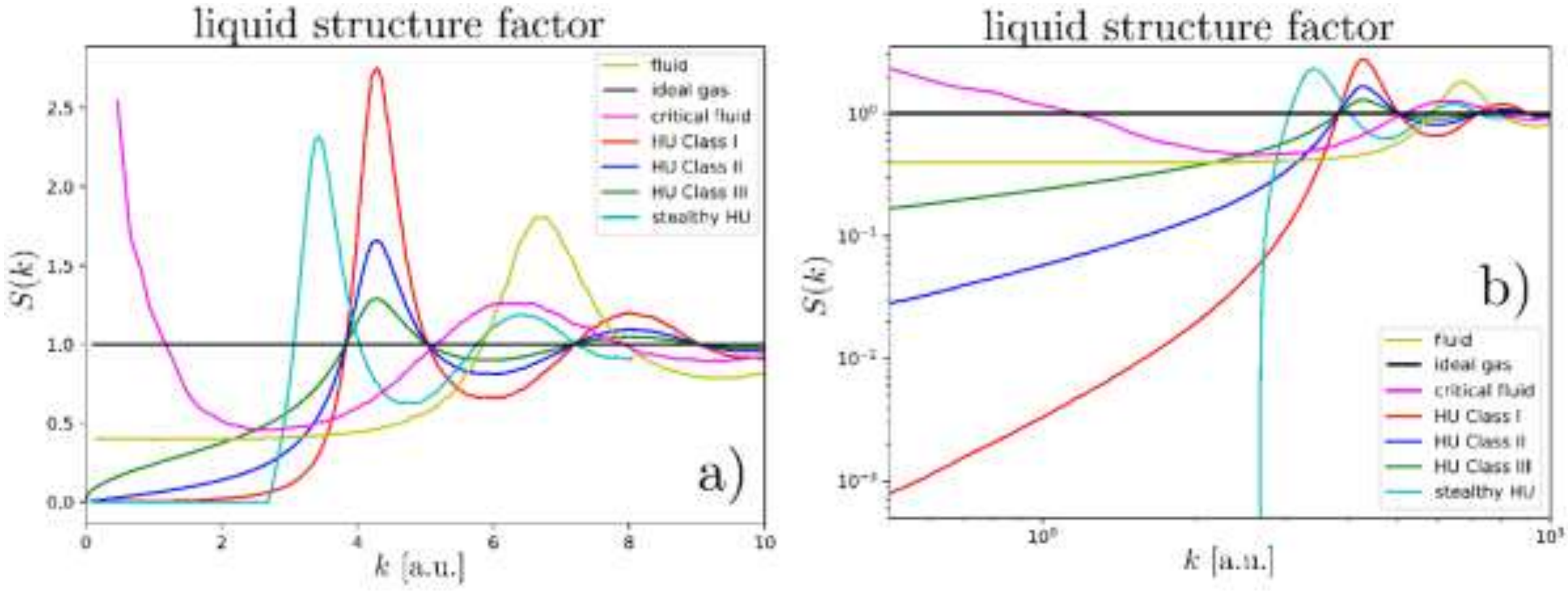

Figure 1: *Schematics of liquid structure factors.* Static structure factor *S(k)* as a function of the wave number k for a typical strongly interacting liquid (green-yellow), the ideal gas (black), a critical fluid (purple), and a hyperuniform system of Class I (red), Class II (dark blue), and Class III (green), as well as stealthy (light blue). a) Linear scale, b) the same data on a double-logarithmic scale.

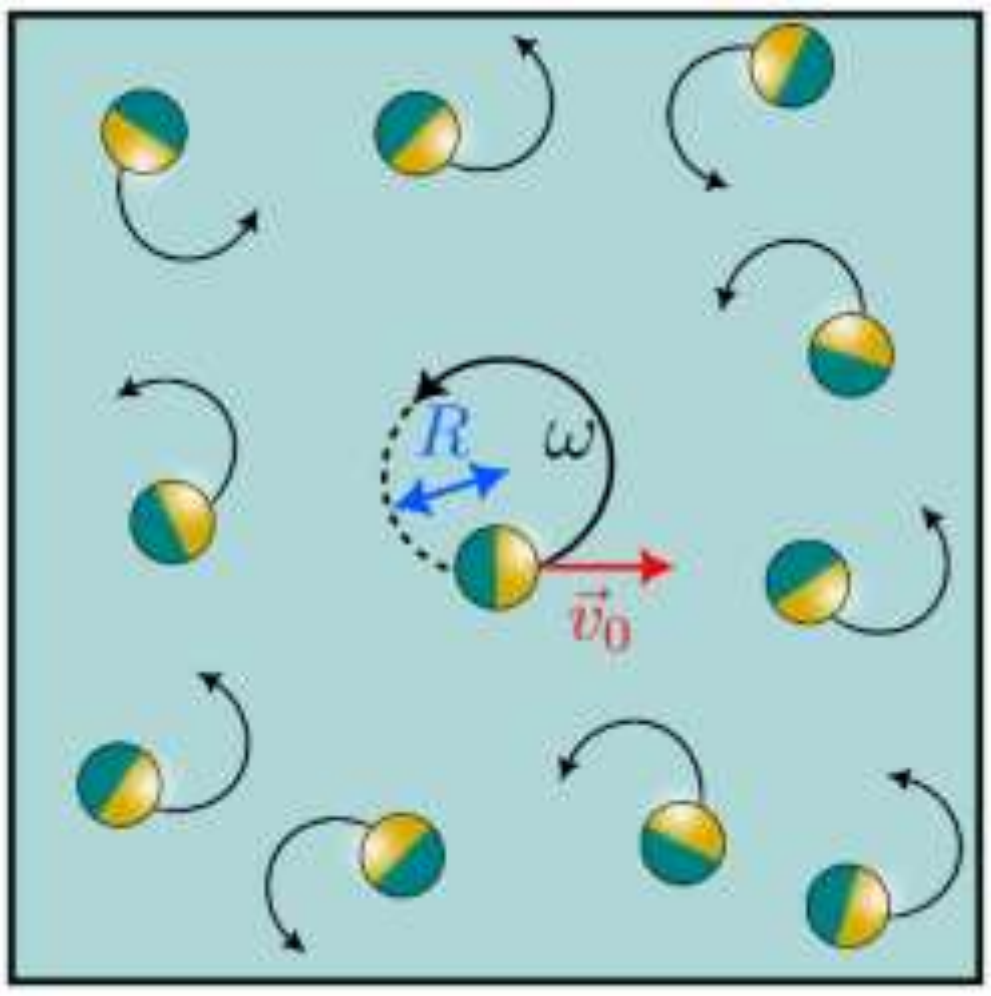

Figure 2: *Sketch of the model of interacting circle swimmers moving in two dimensions.* Individual gold-capped Janus particles move counterclockwise along circles of radius $R$ (blue) with spinning frequency ω and self-propulsion speed $v_0$ (red arrow), in the absence of thermal fluctuations. The curved arrows indicate the ideal trajectories that non-interacting particles would follow. Mutual collisions can lead to disordered hyperuniform configurations.

## Acknowledgements

We thank Christian Bair and Maxim Root for help in preparing the figures. HL was financially supported by the Science Priority Program SPP 2265 of the German Research Foundation (DFG) within project LO 418/25. ECO acknowledges financial support from the National Natural Science Foundation of China under Grant No. 12361141816.